\documentclass[twocolumn,showpacs,nofootinbib]{revtex4}
\usepackage{graphicx}
\usepackage{epsf}
\usepackage{bm}
\usepackage{amsmath}
\usepackage{amsfonts}
\usepackage{amssymb}
\newcommand{\be}{\begin{equation}}
\newcommand{\ee}{\end{equation}}

\newcommand{\mincir}{\raise
-3.truept\hbox{\rlap{\hbox{$\sim$}}\raise4.truept\hbox{$<$}\ }}
\newcommand{\magcir}{\raise
-3.truept\hbox{\rlap{\hbox{$\sim$}}\raise4.truept\hbox{$>$}\ }}

\begin{document}


\title{New cosmic accelerating scenario without dark energy}

\vspace{0.5cm}

\author{J. A. S. Lima$^{1}$\footnote{E-mail: limajas@astro.iag.usp.br}}

\author{S. Basilakos$^{2}$\footnote{E-mail: svasil@Academyofathens.gr}}

\author{F. E. M. Costa$^1$\footnote{E-mail: ernandesmc@usp.br}}

\vspace{0.8cm}

\affiliation{$^1$Departamento de Astronomia, Universidade de S\~ao Paulo, 55080-900, S\~ao Paulo, SP, Brazil}

\affiliation{$^2$Academy of Athens, Research Center for Astronomy and Applied
Mathematics, Soranou Efesiou 4, 11527, Athens, Greece}


\date{\today}

\begin{abstract}
We propose an alternative, 
nonsingular, cosmic scenario based on gravitationally induced particle 
production. The model is an attempt to evade the
coincidence and cosmological constant problems of the standard model ($\Lambda$CDM) and also 
to connect the early and late time accelerating stages of the Universe. 
Our space-time emerges from a pure initial de Sitter stage thereby 
providing a natural solution to the horizon problem.  Subsequently, 
due to an instability provoked by the production of massless  particles, the Universe 
evolves smoothly to the standard radiation dominated era thereby 
ending the production of radiation as required 
by the conformal invariance. Next, the radiation 
becomes subdominant with the Universe entering in the cold dark 
matter dominated era. Finally,  the negative pressure associated 
with the creation of cold dark matter (CCDM model) particles accelerates the 
expansion and drives the Universe to a final de Sitter stage.  The 
late time cosmic expansion history of the CCDM model is exactly 
like in the standard 
$\Lambda$CDM model, however, there is no dark energy. The model 
evolves between two limiting (early and late time) de Sitter 
regimes. All the stages are also discussed 
in terms of a scalar field description. This complete  scenario is fully determined by two extreme energy 
densities, or equivalently, the associated de Sitter Hubble scales 
connected by $\rho_I/\rho_f=(H_I/H_f)^{2} \sim 10^{122}$, a result that has no 
correlation with the cosmological constant problem.   
We also study the linear growth of matter perturbations 
at the final accelerating stage. It is found that the 
CCDM growth index can be written 
as a function of the $\Lambda$ growth index, $\gamma_{\Lambda} \simeq 6/11$. 
In this framework, we also compare the observed growth rate of 
clustering with that predicted by the current CCDM model. 
Performing a $\chi^{2}$ statistical test 
we show that the CCDM model provides growth rates that 
match sufficiently well with the 
observed growth rate of structure.

\end{abstract}


\maketitle

\section{Introduction}

The existence of a cosmological constant $\Lambda$ which
can be used in order to explain the recent cosmic acceleration, has
brought the following major theoretical problem: within the
framework of the quantum field theory the vacuum energy density is
more than 120 orders of magnitude larger than the observed $\Lambda$
value measured by the current cosmological data. This is the so
called ``old" cosmological constant problem (CCP) \cite{weinberg89,Lambda}.
The ``new" problem \cite{Peebles03} asks why is the vacuum density
so similar to the matter density just now? Many solutions to both
theoretical problems have been proposed in the literature
\cite{Zlatev99,AASW09,coincidence}. An easy way to overpass the above
problems is to replace the constant vacuum energy with a dark energy
(DE) that evolves with time. However the nature of DE is far from
being understood. Indeed a main caveat of this methodology is the
fact that the majority of the DE models appeared in the literature
are plagued with no physical basis and/or many free parameters.

Nevertheless, there are other possibilities to explain the present
accelerating stage. In particular, the inclusion of the backreaction in the
Einstein Field Equations (EFE) via an effective pressure (which is
negative for an expanding space-time) opened the way for
cosmological applications. In these models, the 
gravitational production of radiation or cold dark matter 
provides a mechanism for cosmic acceleration as earlier 
discussed in \cite{Prigogine,LCW,LG92}.
As a consequence, several interesting
features of cosmologies where the dark sector is reduced due to the
creation of CDM matter have been discussed in the last decade
\cite{LSS08,SSL09,LJO,BL10}. 

In brief, the merits of the particle
creation scenario with respect to the usual DE ideology are a) the
former has a strong physical basis namely nonequilibrium
thermodynamics, while the latter (DE) has not and b) the particle creation
mechanism unifies the dark sector (dark energy and dark matter),
since a single dark component (the dark matter) needs to be
introduced into the cosmic fluid and thus it contains only one free
parameter. Interestingly, from the viewpoint of a statistical  Bayesian analysis models
which include only one free parameter should be preferred along the
hierarchy of cosmological models \cite{GL}. We would like to emphasize here
that the only cosmological models (to our knowledge) which satisfy
the above statistical condition are:

i) the concordance $\Lambda$CDM
which however suffers from the coincidence and fine tuning problems \cite{Zlatev99,AASW09,coincidence}.

ii) the braneworld cosmology of \cite{Deff} which however does not
fit the SNIa+BAO+CMB(shift-parameter) data (see \cite{BasPLim10}).

(iii) the current particle creation
model which simultaneously fits the observational data and 
alleviates the coincidence and fine tuning problems \cite{LJO,BL10}.

In this paper, we are proposing a new cosmological scenario which is complete in the following sense: 
all the accelerating stages of the cosmic evolution are powered
uniquely by the gravitational creation of massless (at the very early stage) and massive cold dark matter
particles (at the late stages).

In our scenario, the Universe starts from an unstable de
Sitter dominated phase ($a \propto e^{H_I t}$) powered by 
the production of massless
particles, and, as such, there is no the horizon problem.
Subsequently, it deflates and evolves to the standard
radiation phase ($a \propto t^{1/2}$) thereby ending 
the creation of massless particles.
Due to expansion, the radiation becomes subdominant with the
Universe entering in the cold dark matter (CDM) 
dominated era, in which the linear growth of matter fluctuations is taking place in a natural way.
Finally, the negative pressure associated with the creation of cold
dark matter particles accelerates the expansion and drives the
Universe to a final de Sitter stage. In addition, the
transition from Einstein-de Sitter ($a \propto t^{2/3}$) to a de Sitter final stage ($a \propto e^{H_f t}$)
guarantees the consistence of the model with the supernovae type Ia
data and complementary observations, including the growth rate 
of clustering.  A transition redshift of the
order of a few (exactly the same value predicted by $\Lambda$CDM) is
also obtained.

The paper is structured as follows. In Sec. II, we discuss 
the basic ideas underlying the 
particle production in an 
expanding Universe and set up the basic equations whose solutions describe 
the complete evolution of our model.  In Sec. III, we 
study the linear growth of perturbations, 
whereas in Sec. IV, we constrain the growth index through 
a statistical analysis involving the 
latest observational results. In Sec. V, a possible scalar 
field description for all stages 
is discussed, and, finally, in Sec. VI we summarize the basic results.

\section{A Complete Cosmological Scenario with Gravitational Particle Production}
The microscopic description for gravitationally induced particle
production in an expanding Universe began with Schr\"odinger's
\cite{Scho39} seminal paper, which referred to it as an alarming
phenomenon. In the late 1960s, this issue was rediscussed by Parker
and others \cite{Parker,BDbook,Muka} based on the Bogoliubov mode-mixing
technique in the context of quantum field theory in a curved
space-time described by the general relativity theory \cite{FR}. Physically, one may think that the (classical) time
varying gravitational field works like a `pump' supplying energy to
the quantum fields.

In order to understand the basic approach,  let us consider a
real minimally coupled massive scalar field $\phi$ evolving in a
flat expanding Friedman-Robertson-Walker (FRW) geometry. In units
where $\hbar=k_B=c=1$, the field is described by the following action

\begin{equation}\label{m63}
S={1\over 2} \int \sqrt{-g}d^4 x
\bigg[g^{\alpha\beta}\partial_\alpha\phi\partial_\beta\phi-m^2
\phi^2\bigg]\,.
\end{equation}
In terms of the conformal time $\eta$ ($dt = a(\eta)d\eta$), the
metric tensor $g_{\mu\nu}$ is conformally equivalent to the
Minkowski metric $\eta_{\mu\nu}$, so that the line element is
$ds^2=a^2(\eta)\eta_{\mu\nu}dx^\mu dx^\nu$, where $a(\eta)$ is the
cosmological scale factor. Writing the field $\phi (\eta,x) =
a(\eta)^{-1}\chi$, one obtains from the above action
\begin{equation}\label{m67}
\chi''- \nabla^2 \chi +\bigg( m^2a^2-{a''\over a}\bigg)\chi=0\,,
\end{equation}
where the prime here denotes derivatives with respect to $\eta$. Notice
that the field $\chi$ obeys the same equation of motion as a massive
scalar field in Minkowski space-time, but now with a time dependent
{\em effective mass},
\begin{equation}\label{m68}
m^2_{eff}(\eta)\equiv m^2a^2-{a''\over a}\,.
\end{equation}
This time varying mass accounts for the interaction between the
scalar and the gravitational fields.  The energy  of the field
$\chi$ is not conserved (its action is explicitly time time dependent),
and, more important, its quantization leads to particle creation at
the expense of the classical gravitational background
\cite{Parker,BDbook,Muka}.

On the other hand, in the framework of general relativity theory,
the scale factor of a FRW type universe dominated by
radiation ($a \propto t^{1/2}$) satisfies the following relation $[a{\ddot a} + {\dot a}^{2} = 0]$,
or, in the conformal time, $a'' = 0$. Therefore, for  massless
fields ($m=0$), there is no particle production  since Eq.
(\ref{m67}) reduces to the same of a massless field in Minkowski
space-time, and, as such,  its quantization becomes trivial. This is the basis of
Parker theorem concerning the absence of massless particle
production in the early stages of the Universe. Note that Parker's
result does not forbid the production of massless particles in a
very early de Sitter stage ($a''\neq 0$). Potentially, we also see  
that massive particles
can always be produced by a time varying gravitational field. 
As we shall see, such features are
incorporated in the scenario proposed here.

In principle, for applications in cosmology, the above semiclassical
results have three basic difficulties, namely:

(i) The scalar field was treated as a test field, and, therefore,
the FRW background is not modified by the newly produced particles.

(ii) The particle production is an irreversible process, and, as
such, it should be constrained by the second law of thermodynamics.

(iii) There is no a clear prescription of how an irreversible
mechanism of quantum origin can be incorporated in the EFE.

Later on, a possible macroscopic solution for these problems was put
forward by Prigogine and coworkers \cite{Prigogine} using
nonequilibrium thermodynamics for open systems, and by  Calv\~ao,
Lima \& Waga \cite{LCW}  through a covariant relativistic treatment
for imperfect fluids (see also \cite{LG92}). The novelty of such an
approach is that particle production, at the expense of the
gravitational field, is an irreversible process constrained by the
usual requirements of nonequilibrium thermodynamics.  This
irreversible process is described by a negative pressure term in the
stress tensor whose form is constrained by the second law of
thermodynamics\footnote{The quantum approach is unable to
provide the entropy burst accompanying the particle production since
it is adiabatic and reversible.}. This macroscopic
approach has also microscopically been justified by Zimdahl and
collaborators through a relativistic kinetic theoretical formulation
\cite{ZP94}. In comparison to the standard equilibrium equations,
the irreversible creation process is described by two new
ingredients: a balance equation for the particle number density and
a negative pressure term in the stress tensor. Such quantities are
related to each other in a very definite way by the second law of
thermodynamics. Since the middle of the nineties, several
interesting features of cosmologies with creation of cold dark
matter  and radiation have been investigated by many authors
\cite{LSS08,ZP2,ZP4,ZIM00,ZP5}.

In what follows, as theoretically predicted by inflation and observationally indicated the angular power spectrum 
of the temperature fluctuations, we  consider the EFE for a homogeneous, isotropic, 
spatially flat universe with gravitationally induced particle production:
\begin{equation}\label{friedr}
8\pi G\rho = 3 \frac{\dot{a}^2}{a^2},
\end{equation}
\begin{equation}\label{friedp}
8\pi G(p + p_{c})=  -2 \frac{\ddot{a}}{a} -
\frac{\dot{a}^2}{a^2},
\end{equation}
where an overdot means time derivative, $\rho$ and p are the
dominant energy density and pressure of the cosmic fluid,
respectively, and $p_c$ is a dynamic pressure which depends on the
particle production rate. Special attention has been paid to the
simpler process termed  ``adiabatic" particle production. It means
that particles and entropy are produced in the space-time, but the
specific entropy (per particle), $\sigma = S/N$, remains constant
\cite{LCW}. In this case,  the creation pressure reads \cite{Prigogine,LCW,LG92,LSS08,SSL09}            
\begin{equation}\label{CP}
p_{c} = -\frac{(\rho + p) \Gamma}{3H},
\end{equation}
where $\Gamma$ with dimensions of $(time)^{-1}$ is the particle
production rate and $H={\dot a}/a$ is the Hubble parameter \cite{2nd}.  In principle, the quantity $\Gamma$ should be determined 
from quantum field theory in curved spacetimes by taking into account that particle production is an irreversible process.  

But how the evolution of $a(t)$ affected by $\Gamma$? By
assuming a dominant cosmic fluid satisfying the equation of state
(EoS), $p=\omega\rho$, where $\omega$ is a constant, the EFE imply
that
\begin{equation}
\label{varH0} \dot H + \frac{3}{2} (1 + \omega)  H^2
\left(1-\frac{\Gamma}{3H}\right) = 0.
\end{equation}
The de Sitter solution ($\dot H=0$,  $\Gamma = 3H = constant$) is
now possible regardless of the  EoS defining the cosmic fluid. Since
the Universe is evolving, such a solution is unstable, and, as long
as $\Gamma \ll 3H$, conventional solutions without particle
production are recovered. From the above equation, one may conclude that the main effect of $\Gamma$ is to provoke a
dynamic instability in the space-time thereby allowing a transition
from a de Sitter regime ($\Gamma \sim 3H$) to a conventional solution, and vice versa (see subsections A and B below).


\subsection{From an early de Sitter stage to the standard radiation phase}

Let us first discuss the transition from an initial de Sitter stage
to the standard radiation phase. The main theoretical constraints 
are:

\begin{itemize}

 \item{\it The model must not only solve the horizon problem  but also provide a
 quasiclassical boundary condition to quantum cosmology (a hint on how to solve the initial singularity problem).}

\item{\it Massless particles  cannot be quantum-mechanically produced in the conventional radiation phase (Parker's Theorem).}

\end{itemize}
To begin with, let us  assume a radiation dominated Universe
($\omega=1/3,\, \Gamma\equiv\Gamma_r $). The dynamics is determined
by the ratio $\Gamma_r/3H$ [see Eq. (\ref{varH0})].  The most natural
choice would be a ratio which favors no epoch in the evolution of
the Universe ($\Gamma_r/3H = constant$). However, the particle
production must be strongly suppressed, $\Gamma_r/3H \ll 1$, when the
Universe enters the radiation phase. 
The simplest formula satisfying
such a criterion is linear, namely: 
\begin{equation}
\label{evarH1} 
\frac{\Gamma_r}{3H} = \frac{H}{H_I},
\end{equation} 
where $H_I$ is the inflationary  expansion rate 
associated to the initial de Sitter ($H \leq H_I$).
It is worth to notice that such particle creation rates   
have been previously discussed by several 
authors (see \cite{Prigogine,LG92,LA99,ZIM00} and references therein). 
It is also worth noticing that for ``adiabatic'' photon creation the form of the blackbody spectrum is preserved in the course of the expansion \cite{Spect96}.

Now, inserting Eq.(\ref{evarH1}) into Eq. (\ref{varH0}) it becomes:

\begin{equation}
\label{varH1} \dot H + {2}  H^2 \left(1-\frac{H}{H_I}\right) = 0.
\end{equation}

The solution of the above equation  can be written as

\begin{equation}
\label{solH} H(a) = \frac{H_I}{1+ D a^{{2}}},
\end{equation}
where $D \geq 0$ is an integration constant. Note that $H=H_I$ is a
special solution of Eq. (\ref{varH1}) describing the exponentially
expanding de Sitter space-time. This solution is  unstable with
respect to the critical value $D=0$. For $D>0$, the universe starts
without a singularity and evolves continuously towards a radiation
stage, $a \sim {t}^{1/2}$, when $Da^{2} \gg 1$. By integrating
Eq. (\ref{solH}), we obtain the scale factor:

\begin{equation}
H_I t=\ln \frac{a}{a_*}+\frac{\lambda^{2}}{2}{\left(\frac{a}{a_{*}}\right)}^{2},
\label{full1}
\end{equation}
where $\lambda^{2}=Da_*^{2}$ is an integration constant and $a_*$
defines the transition from the de Sitter stage to the beginning of
the standard radiation epoch\footnote{This kind of evolution was
first discussed by G. L. Murphy \cite{Murphy73} by studying possible
effects of the second viscosity in the very early Universe. Later
on, it was also investigated in a more general framework involving
cosmic strings by J. D. Barrow \cite{Barrow88} who coined the
expression ``Deflationary Universes".  It has also been discussed in
connection with decaying $\Lambda(t)$-models \cite{LT96}.}. At early
times ($a \ll a_{*}$), when the logarithmic term dominates, one
finds $ a \simeq  a_*e^{H_I t}, $ while at late times, $a \gg a_*, H
\ll H_I$, Eq. (\ref{full1}) reduces to $a \simeq a_*
\left(\frac{2H_I}{\lambda^{2}}t\right)^{1/2}$, and the  standard radiation phase is reached.

It should be noticed that the time scale ${H_I}^{-1}$ provides the greatest value of the
energy density, $\rho_{I}=3H_{I}^{2}/8{\pi}G$,
characterizing the initial de Sitter stage which is supported by the
maximal radiation production rate, $\Gamma_r=3H_{I}$.  From Eqs. (\ref{friedr}) and
(\ref{solH}) we obtain the radiation energy density:

\begin{equation}\label{rho1}
\rho_r={\rho_{I}}\left[1 +
\lambda^{2}\left(\frac{a}{a_*}\right)^{2}\right]^{-2}.
\end{equation}
As expected, we see again that the conventional radiation phase, $\rho_r \sim a^{-4}$,
is attained when $a \gg a_*$.

Now we pose the following question: how does the cosmic 
temperature evolve at the very early stages? For ``adiabatic" production of relativistic particles
the energy density scales as $\rho_{r} \sim T^{4}$
\cite{LCW,LG92}, and the above equation implies that

\begin{equation}\label{T}
T={T_{I}}\left[1 +
\lambda^{2}\left(\frac{a}{a_*}\right)^{2}\right]^{-1/2},
\end{equation}
where $T_I$ is the temperature of the initial de Sitter phase which
must be uniquely determined by the scale $H_I$. We see that the
expansion proceeds isothermally during the de Sitter phase
($a\ll a_*$) which means that the supercooling and subsequent 
reheating that is taking
place in several inflationary variants are avoided \cite{Turner83,KT90}.
In other words, there is no 
`graceful exit' problem.


After the de Sitter stage, the temperature decreases continuously in the
course of the expansion. For $a\gg a_*$ ($H\ll H_I$), we obtain $T \sim
a^{-1}$. Accordingly, the comoving number of photons becomes
constant since $n \propto {a}^{-3}$, as expected for the standard
radiation stage.\footnote{Since $n_r \propto T^{3}$,  the average
photon concentration reads $n_r=n_{I}\left[1 +
\lambda^{2}({a}/{a_*})^{2}\right]^{-{3}/{2}}$.}

In this context, we also need to answer to the following 
question: what about the initial temperature $T_I$? 
Since the model
starts as a de Sitter space-time, the most natural choice is to
define $T_I$ as the Gibbons-Hawking temperature \cite{GibHaw} of its
event horizon, $T_I=H_{I}/2\pi$.
Naively, one may expect $T_I$ of the same order or smaller than the
Planck temperature because of the classical description. From EFE we
have $\rho_I = 3{m_{Pl}}^{2}{H_I}^{2}/8 \pi$ (where $m_{Pl}\simeq
1.22\times 10^{19} GeV$), and since the energy density is $\rho_I =
N_*(T)T_I^{4}$, one finds  $T_I \sim H_I \sim 10^{19} GeV$ (where
$N_* (T) = {\pi}^2 g_*(T)/30$ depends on the number of effectively
massless particles).

Naturally, due to the initial de Sitter phase, the model is free of particle horizons. A light pulse beginning at
$t=-\infty$ will have traveled by the cosmic time $t$ a physical
distance, $d_{H}(t)=
a(t)\int_{-\infty}^{t}\frac{d\tilde{t}}{a(\tilde{t})}$, which
diverges thereby implying the absence of 
particle horizons. The latter feature means that the local 
interactions may homogenize the whole Universe.


Since photons are not produced in the radiation phase, the 
Big-Bang Nucleosynthesis (BBN) may work in
the conventional way \cite{Steigman}.  Subsequently, the Universe 
enters the cold
dark matter [Einstein-de Sitter, $a(t)\propto t^{2/3}$] dominated phase.
Finally,  we have also verified that  a large class of 
$\Gamma_{r}$ is capable to overcome  the  ``graceful exit'' and 
fine tuning problem. 
Indeed we have found that the correct  transition from an early 
de Sitter to the radiation  phase is valid even for 
$\Gamma_{r} \propto H^{n}$ with $n \geq 2$ (work in progress). 
The details are of course different depending on the power $n$, but 
the qualitative fact of the
transition is universal, and to our opinion this is very good news
because it shows that it can be a clue for a general graceful exit
mechanism.

\subsection{From Einstein-de Sitter to a late time de Sitter stage}

Due to the conservation of baryon number the remaining question is the
production rate of cold dark matter particles and the overall late
time evolution.  In other words, what is the form of $\Gamma_{dm}$?
For simplicity, we consider here only the dominant CDM component.

In principle, $\Gamma_{dm}$ should be determined  from quantum field
theory in curved spacetimes.  In the absence of a rigorous
treatment, we consider  (phenomenologically)  the  following fact
\cite{K08,Peebles,Kom11}:

\begin{figure}[ht]
\includegraphics[width=0.5\textwidth]{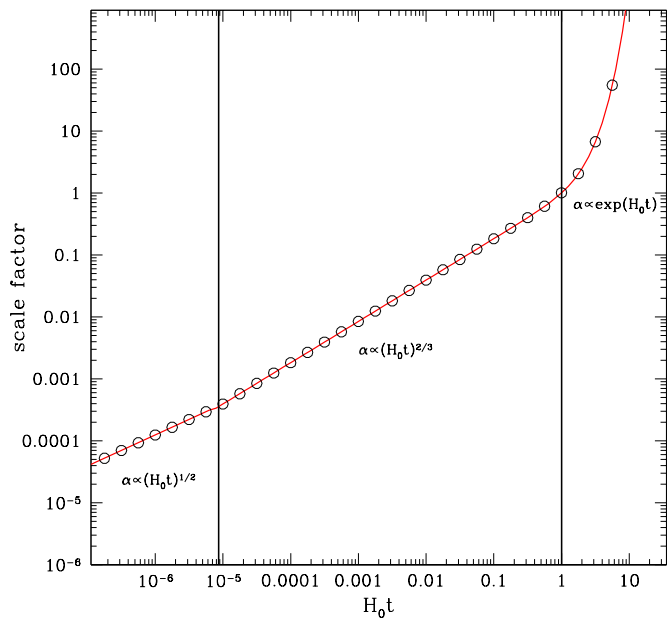}
\caption{Overall evolution
of the scale factor predicted by the matter creation model (solid
line) and the traditional $\Lambda$CDM cosmology (open points). In
this plot we have adopted the best fit,
$\tilde{\Omega}_{m}=0.273$, from WMAP7 \cite{Kom11}.}
\end{figure}

\begin{itemize}
\item{\it All available observations  are in accordance with the $\Lambda$CDM evolution both at the
background and perturbative levels.}
\end{itemize}
Now, we recall that a flat $\Lambda$CDM model evolves like:
\begin{equation} \label{LCDM}
\dot H + \frac{3}{2} H^2 \left[1 -
\left(\frac{H_f}{H}\right)^{2}\right]=0,
\end{equation}
where ${H_f}^{2} = \Lambda/3$ sets the Hubble scale of the final de
Sitter stage ($H \geq H_f$). Such  behavior should be compared to
that predicted for a dust filled model ($\omega = 0,\,
\Gamma\equiv\Gamma_{dm}$) with particle production [see Eq.
(\ref{varH0})]:
\begin{equation}
\label{varCDM} \dot H + \frac{3}{2} H^2 \left(1-
\frac{\Gamma_{dm}}{3H} \right) = 0.
\end{equation}
By comparing Eqs. (\ref{LCDM}) and (\ref{varCDM}), we see that the same
background evolution requires that $\Gamma_{dm}/{3H}=
\left({H_f}/{H}\right)^{2}$. 
Thus, in the background solution, the particle creation 
rate does not depend on a 
given scale but rather it is considered homogeneous.
The limiting value of the creation rate, $\Gamma_{dm}=3H_f$, leads
to a late time de Sitter phase ($\dot H=0,\, H=H_f$) thereby showing
that the de Sitter solution now becomes  an attractor at late times.
With this proviso, the solution of Eq. (\ref{varCDM}) reads:


\begin{equation}
\label{Hz}
 {H}^{2} = {H_0}^2{E(z)}^2={H_0}^2 \left[
\tilde{\Omega}_{m}(1+z)^3+\tilde{\Omega}_{\Lambda} \right],
\end{equation}
where $\tilde{\Omega}_{\Lambda} \equiv (H_f/H_0)^{2}=1-\tilde{\Omega}_{m}$ 
is smaller than unity and $1+z=a^{-1}$. Such solution mimics the Hubble
function $H(z)$ of the traditional flat $\Lambda$-cosmology, with
$\tilde{\Omega}_{\Lambda}$ playing the dynamical role of
$\Omega_{\Lambda}$ (dark energy appearing in 
the concordance model)\footnote{Observationally there is no reason 
to distinguish between $\tilde{\Omega}_{\Lambda}$ which appears 
in Eq. (\ref{Hz}) and that of the cosmological constant term 
parametrized by $\Omega_{\Lambda}$.}.
The dark matter parameter ($\Omega_{dm}=1$) is also replaced by an
effective parameter, $({\Omega_{dm}})_{eff} \equiv
1-\tilde{\Omega}_{\Lambda}$, which quantifies the amount of matter
that is  clustering. This explains why this model is in agreement
with the dynamical determinations related to the amount of the cold
dark matter at the  cluster scale, and, simultaneously, may also be
compatible with the position of the first acoustic peak in the
pattern of CMB anisotropies which requires $\Omega_{total}=1$.

By integrating  Eq. (\ref{Hz}) we obtain

\begin{equation}
a(t)=\left(\frac{\tilde{\Omega}_{m}}{\tilde{\Omega}_{\Lambda}}\right)^{1/3}
\sinh^{\frac{2}{3}}\left(\frac{3H_{0}\sqrt{\tilde{\Omega}_{\Lambda}}
}{2}t\right).
\end{equation}

Note that the late time dynamics is determined by a single
parameter, namely: $\tilde{\Omega}_{\Lambda}=1-\tilde{\Omega}_{m}$, 
and is identical to that predicted by the flat $\Lambda$CDM model.
Using the current, a joint
statistical analysis, involving the latest observational data
(SNIa~\cite{Ama11}, BAO~\cite{Perc10} and CMB shift parameter~\cite{Kom11})
is implemented. We find that the overall likelihood function peaks at 
${\tilde \Omega}_{m}=0.274\pm 0.011$ with 
$\chi_{\rm tot}^{2}({\tilde \Omega}_{m})
\simeq 543.18$ for $557$ degrees of freedom. Since the current statistical 
results are in excellent agreement with those provided by 
WMAP7 \cite{Kom11}, for the rest of the paper we will 
restrict our present analysis
to the choice $({\tilde \Omega}_{m},\sigma_{8,0})=(0.273,0.811)$,
where $\sigma_{8,0}$ is the rms mass fluctuations on scales of 
$8h^{-1}$ Mpc at redshift $z=0$.

In Figure 1, we show the overall evolution 
(radiation, matter and dark energy dominated eras)
of our complete cosmological scenario which coincides exactly
with the one recently discussed in Refs. \cite{LJO,BL10} following a slightly
different approach.
Note also that by replacing
the value of $\Gamma_{dm}$ into the definition of the creation
pressure [see Eq. (\ref{CP})] one obtains that it is negative and
constant ($p_c = - 3H_f^{2}/ 8\pi G =
-3\tilde{\Omega}_{\Lambda}H_0^{2}/8\pi G$).


\section{The Evolution of the linear growth factor}
In this section, we briefly discuss the basic equation which 
governs the behavior of the linear matter
perturbations $\delta_{m}\equiv \delta\rho_m/\rho_m$
on subhorizon scales in the CCDM model, assuming that 
the particle creation rate remains homogeneous 
and only the corresponding effective dark matter
forms structures. The reason for introducing the 
growth analysis here is to give the reader the 
opportunity to appreciate also at the perturbative level, 
the relative strength and similarities
of the CCDM and $\Lambda$CDM models used to constrain the growth index. 
As discussed 
by Jesus {\it et al.} \cite{Jesus10} based 
on the Neo-Newtonian approach \cite{LZB},  the 
evolution equation of
the matter fluctuations of a  CCDM cosmology  reads:

\begin{equation}\label{odedelta}
\frac{d^{2}\delta_{m}}{d\eta^{2}}+F(\eta)
\frac{d\delta_{m}}{d\eta}+{\cal G}(\eta)\delta_{m}=0,
\end{equation}
whose solution is  
$\delta_{m}(\eta) \propto {\cal D}(\eta)$, with ${\cal D}(\eta)$ denoting the 
linear growing mode
(usually scaled to unity at the present time). The functions appearing in (\ref{odedelta}) are defined by:
\begin{equation}\label{FCef}
F(\eta) =  \frac{\tilde{\Omega}_{m}
(1+6c_{p})+2\tilde{\Omega}_{\Lambda} e^{3\eta}(8+3c_{p})}
{2(\tilde{\Omega}_{m} +\tilde{\Omega}_{\Lambda} e^{3\eta})}\;,
\end{equation}
\begin{eqnarray}
{\cal G}(\eta) &=& \frac{9\tilde{\Omega}^{2}_{m}}
{2(\tilde{\Omega}_{m} +\tilde{\Omega}_{\Lambda}e^{3\eta})^2}\nonumber\\
&+& \frac{15\tilde{\Omega}_{\Lambda}
e^{3\eta}(1+c_{p})-3\tilde{\Omega}_{m}
(2+c_{p})}{\tilde{\Omega}_{m} +\tilde{\Omega}_{\Lambda}e^{3\eta}}\;,
\end{eqnarray}
where $\eta={\rm ln}a(t)$ and $\tilde{\Omega}_{m}=1-\tilde{\Omega}_{\Lambda}$. 
The quantity $c_{p}$ can be viewed as 
the ``effective adiabatic'' sound speed
\begin{equation}
c_{p}\equiv c^{2}_{eff,ad}=\frac{\delta p_{c}}{\delta \rho_{m}}\;,
\end{equation}
where $\delta p_{c}$ is the perturbation of creation pressure.
Obviously using the above equation and Eq.(\ref{CP}) we find that 
the corresponding ``effective adiabatic'' sound speed must be negative
$c_{p}=c^{2}_{eff,ad}<0$. Note that the 
latter restriction is valid also for the interacting dark energy models 
(see \cite{Valiviita2008})\footnote{Based on a scalar field 
description (in our case see section V) one can show
that the sound speed 
$c^{2}_{s \phi}=\delta p_{\phi}/\delta \rho_{\phi}$ is exactly unity.
Indeed using similar arguments to those of \cite{Valiviita2008}
we also prove that     
$\delta \rho_{\phi}=\delta(\frac{{\dot \phi}^{2}}{2}+V)=
{\dot \phi}\delta {\dot \phi}=\delta(\frac{{\dot \phi}^{2}}{2}-V)=\delta p_{\phi}$ which implies $c^{2}_{s \phi}=1$.
Valiviita {\it et al.} \cite{Valiviita2008} 
proposed that in order to avoid instabilities in the
dark energy we have to have $c^{2}_{s \phi}=1$ and $c^{2}_{eff,ad}<0$. 
Obviously both conditions are full-filled here.}.
We would like to stress that for simplicity we are using
a constant $c_{p}$ (for more discussions see 
\cite{Jesus10}). In this context, the functions $F(\eta)$ and ${\cal G}(\eta)$ 
defined above can be written in terms of $\Omega_{m}(\eta)$ as
\begin{equation}\label{F11}
F(\Omega_{m})=\frac{1+c_{p}+15(1-\Omega_{m})}{2}\;,
\end{equation}
and
\begin{equation}\label{G11}
{\cal G}(\Omega_{m})=\frac{9}{2}\Omega^{2}_{m}+15(1+c_{p})(1-\Omega_{m})
-3(2+c_{p})\Omega_{m}\;, 
\end{equation}
where  the  definition
\begin{equation}\label{OF11}
\Omega_{m}(\eta)=1-\Omega_{\Lambda}(\eta)=\frac{{\tilde \Omega}_{m}}
{{\tilde \Omega}_{m} +{\tilde \Omega}_{\Lambda}e^{3\eta} },
\end{equation}
has been  adopted. At this point, we remind the reader that solving 
Eq.(\ref{odedelta}) for
the  $\Lambda$CDM cosmology\footnote{For the usual 
$\Lambda$CDM cosmological model we have 
$\frac{df}{d\Omega_{m}}\frac{d\Omega_{m}}{d\eta}+f^{2}+X(\eta)f+{\cal G}(\eta)=0$
where 
$X(\eta)=\frac{1}{2}+\frac{3}{2}\left[1-\Omega_{m}(\eta)\right]$ and
${\cal G}(\eta)=-\frac{3}{2}\Omega_{m}(\eta)$.}, we derive
the well-known perturbation growth factor (see \cite{Peeb93}):
\begin{equation}
\label{eq24}
{\cal D}_{\Lambda}(z)=\frac{5{\tilde \Omega}_{m}
  E(z)}{2}\int^{+\infty}_{z}
\frac{(1+u)du}{E^{3}(u)} \;\;.
\end{equation}
Obviously, for $E(z) \simeq {\tilde \Omega}_{m}^{1/2}\,(1+z)^{3/2}$ it
gives the standard result ${\cal D}(z)\simeq a=e^{\eta}=(1+z)^{-1}$, which
corresponds to the matter dominated epoch, as expected.

Now, for any type of DE, an efficient parametrization
of the matter perturbations is based on the growth rate of clustering
originally introduced by Peebles \cite{Peeb93}. This is
\begin{equation}
\label{fzz221}
f(\eta)=\frac{d\ln \delta_{m}}{d\eta}\simeq \Omega^{\gamma}_{m}(\eta),
\end{equation}
which implies
\begin{equation}
\label{eq244}
{\cal D}(a)={\rm exp} \left[\int_{1}^{a}
\frac{\Omega_{m}^{\gamma}(x)}{x} dx \right],
\end{equation}
where $\gamma$ is the so called growth index
(see Refs. \cite{Silv,Wang98,Linjen03,Lue04,Linder2007,Nes08})
which plays a key role in cosmological studies, especially in the light of
recent large redshift surveys 
(like the {\em WiggleZ} and SDSS (DR9); see \cite{Blake,Sam11,Reid12} 
and references therein). 
As an example, it was theoretically shown that for DE models
which adhere to general relativity the growth index $\gamma$ is
well approximated in terms of the equation of state parameter
$\gamma_{DE} \simeq \frac{3(w-1)}{6w-5}$
(see \cite{Silv,Wang98,Linder2007,Nes08}), which
boils down to $\gamma_{\Lambda}\approx 6/11$ for
the $\Lambda$CDM cosmology $w(z)=-1$.
Notice, that in the case of the
braneworld model of Dvali, Gabadadze \& Porrati \cite{DGP}
we have $\gamma_{DGP} \approx 11/16$ (see also \cite{Linder2007,Fu09}), while 
for the $f(R)$ gravity models we have $\gamma_{R} \simeq 0.41-0.43$ 
\cite{Gann09} at the present time.

\begin{table*}
\caption[]{Summary of the observed growth rate and references.}
\tabcolsep 6pt
\begin{tabular}{ccccc}
\hline
Index & $z$ & growth rate $(A_{obs})$& Refs.& Symbols (in Fig. 3)\\ \hline \hline
1&0.17 & $0.510\pm 0.060$& \cite{Song09, Perc04}& open circles\\
2&0.35 & $0.440\pm 0.050$& \cite{Song09, Teg06} &open circles\\
3&0.77 & $0.490\pm 0.180$& \cite{Song09, Guzzo08}&open circles\\
4&0.25 & $0.351\pm 0.058$&\cite{Sam11}&open triangles\\
5&0.37 & $0.460\pm 0.038$& \cite{Sam11}&open triangles\\
6&0.22 & $0.420\pm 0.070$&\cite{Blake}& solid circles\\
7&0.41 & $0.450\pm 0.040$& \cite{Blake}&solid circles\\
8&0.60 & $0.430\pm 0.040$& \cite{Blake}&solid circles\\
9&0.78 & $0.380\pm 0.040$& \cite{Blake}&solid circles\\
10&0.57 & $0.427\pm 0.066$& \cite{Reid12}&solid square \\
\end{tabular}
\end{table*}

Differentiating Eq. (\ref{fzz221}) with respect to $\eta$ we have
\begin{equation}
\label{ddfz}
\frac{df}{d\eta}+f^{2}=\frac{1}{\delta_{m}}
\frac{d^{2}\delta_{m}}{d\eta^{2}} \;. 
\end{equation}
Using Eqs. (\ref{odedelta}), (\ref{fzz221}) and (\ref{ddfz}), we find after some algebra
\begin{equation}
\label{fzz222}
\frac{df}{d\Omega_{m}}\frac{d\Omega_{m}}{d\eta}+f^{2}+F(\eta)f+{\cal G}(\eta)=0\;,
\end{equation}
where
\begin{equation}\label{domm}
\frac{d\Omega_{m}}{d\eta}=
-3\Omega_{m}(\eta)\left[1-\Omega_{m}(\eta)\right]\;. 
\end{equation}
Inserting the ansatz 
$f\simeq \Omega^{\gamma(\Omega_{m})}_{m}$ into 
Eq. (\ref{fzz222}), using simultaneously Eqs. (\ref{F11}), (\ref{G11}), 
(\ref{OF11}) and 
performing a first order Taylor expansion
around $\Omega_{m}=1$ (for a similar analysis see \cite{Linder2007, Nes08}) 
we find that the asymptotic value of the growth index to the lowest order is 
\begin{equation}\label{gamm}
\gamma \simeq \frac{3(13+12c_{p})}{11+6c_{p}}=
\frac{\gamma_{\Lambda}(13+12c_{p})}{2(1+\gamma_{\Lambda}c_{p})} .
\end{equation}
Inverting the above equation we have
\begin{equation}\label{camm}
c_{p} \simeq \frac{13\gamma_{\Lambda}-2\gamma}{2\gamma_{\Lambda}(\gamma-6)}\;.
\end{equation}
We have checked for various values of $c_{p}$ and ${\tilde \Omega}_{m}$,
that using Eq. (\ref{gamm}) in
Eq. (\ref{eq244}) the latter provides an excellent approximation
to the numerically obtained
form of $D(\eta)$ appears in Eq. (\ref{odedelta}). Indeed the difference 
between the two approaches is less than $0.1-0.2\%$.
Finally, from the above analysis it becomes clear that the possible 
difference between the CCDM and $\Lambda$CDM
predictions is quantified only
at the perturbative level, via the value of $\gamma$, because 
the two cosmological models share the same Hubble parameter as well 
as the same number of free parameters, namely
the dimensionless matter density at the present epoch 
${\tilde \Omega}_{m}$ and the growth index.
In the case of 
$\gamma \equiv\gamma_{\Lambda}\simeq 6/11$ we 
find $c^{\star}_{p} \simeq -1.008$. Hence expanding 
Eq. (\ref{gamm}) around $c^{\star}_{p}$ we can write
\begin{equation}
\gamma\approx \frac{6}{11}+\gamma_{c^{\star}_{p}}(c_{p}-c^{\star}_{p}) \;,
\end{equation}
where
\begin{equation}
\gamma_{c^{\star}_{p}}=\frac{d\gamma}{dc_{p}}(c^{\star}_{p})=\frac{162}
{(11+6c^{\star}_{p})^{2}} \;. 
\end{equation}

\begin{figure}[ht]
\includegraphics[width=0.5\textwidth]{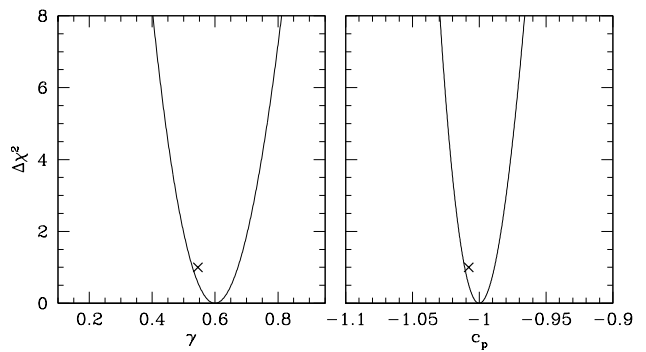}
\caption{{\em Left Panel:}
The variance $\Delta \chi^{2}=\chi^{2}-\chi^{2}_{min}$
around the best fit $\gamma$ value for the CCDM cosmology.
Note that the cross corresponds
to $(\gamma_{\Lambda},\Delta \chi_{1\sigma}^{2})=(\frac{6}{11},1)$.
{\em Right Panel:} The $\Delta \chi^{2}$ versus $c_{p}$. The
corresponding cross is
$(c_{p},\Delta \chi_{1\sigma}^{2})=(-1.008,1)$.}\label{fig:DEE}
\end{figure}

\section{Fitting the CCDM growth index to the Data}

In the following we briefly present some details of the
statistical method and on the observational sample 
that we adopt in order to constrain either 
the growth index or the ``effective adiabatic'' sound speed,
presented in the previous section.

\subsection{The growth data}
The growth data that we will use in this work based on
2dF, VVDS, SDSS and {\em WiggleZ} galaxy surveys,
for which their combination parameter of the growth rate of structure,
$f(z)$, and the redshift-dependent rms fluctuations of the linear
density field, $\sigma_8(z)$,
is available as a function of redshift, $f(z)\sigma_{8}(z)$.
The $f\sigma_{8}\equiv A$ estimator is almost a model-independent
way of expressing
the observed growth history of the universe \cite{Song09}.
In particular the data used are based on

\begin{itemize}
\item The 2dF (Percival et al. \cite{Perc04}), SDSS-LRG 
(Tegmark et al. \cite{Teg06})
and VVDS (Guzzo et al. \cite{Guzzo08})
based growth results as collected by
Song \& Percival \cite{Song09}. This sample contains 3 entries.
\item The SDSS (DR7) results (2 entries)
of Samushia et al. \cite{Sam11}
based on spectroscopic data of $\sim$106000 Luminous Red Galaxies (LRGs)
in the redshift bin $0.16<z<0.44$.
\item The {\it WiggleZ} results of Blake et al. \cite{Blake}
based on spectroscopic data of $\sim$152000 galaxies
in the redshift bin $0.1<z<0.9$. This dataset contains 4 entries.
\item The SDSS (DR9) results of Reid et al. \cite{Reid12}
based on spectroscopic data of $\sim$264000 galaxies
in the redshift bin $0.43<z<0.70$. This dataset includes 1 entry.
\end{itemize}
In Table I we list the precise numerical values of the data points
with the corresponding errors bars.

\subsection{Observational constraints}

In order to constrain the CCDM growth index (or $c_{p}$)
we perform a standard $\chi^2$ minimization procedure between the
growth data measurements (see previous section), 
$A_{obs}=f_{obs}(z)\sigma_{8,obs}(z)$, with the
growth values predicted by the CCDM model at the corresponding redshifts,
$A({\bf p},z)=f({\bf p},z)\sigma_{8}({\bf p},z)$ with
$\sigma_{8}({\bf p},z)=\sigma_{8,0}D({\bf p},z)$.
The vector ${\bf p}$ contains the free parameters of the cosmological model.
In particular,
the essential free parameters entering in the theoretical 

\begin{figure}[ht]
\includegraphics[width=0.5\textwidth]{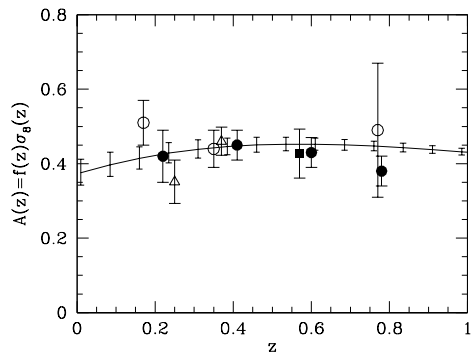}
\caption{Comparison of the observed
and theoretical evolution (see solid line) of the growth
rate of clustering
$A(z)=f(z)\sigma_{8}(z)$.
The thin-line error bars 
correspond to $1\sigma$ $\gamma$ uncertainties. 
The different growth data sets are
represented by different symbols (see Table I for definitions).
}\label{fig:growth}
\end{figure}

\noindent expectation are:
${\bf p} \equiv (\gamma,{\tilde \Omega}_{m})$.
The $\chi^2$ function
is defined as:
\begin{equation}
\label{Likel}
\chi^{2}(z_{i}|{\bf p})=\sum_{i=1}^{N} \left[ \frac{A_{obs}(z_{i})-
A({\bf p},z_{i})}
{\sigma_{i}}\right]^{2}, 
\end{equation}
where $\sigma_{i}$ is the observed growth rate uncertainty.
Note that we sample $\gamma \in [0.1,1.3]$ in steps of 0.001.

In the left panel of Fig. 2 we show the variation of
$\Delta \chi^{2}=\chi^{2}-\chi^{2}_{min}$
around the best $\gamma$ fit value
for the CCDM cosmology.
We find that the likelihood function 
peaks at $\gamma=0.60\pm 0.072$ (or $c_{p}=-1 \pm 0.011$; see the 
right panel of Fig. 2)
with $\chi^{2}_{min} \simeq 7.75$ for $8$ degrees of freedom.
Notice, that for the physically acceptance range 
$0.22\le {\tilde \Omega}_{m} \le 0.32$, we obtain 
either $0.50\le \gamma \le 0.70$ or 
$-1.016 \le c_{p} \le -0.983$
with $\chi^{2}_{min}/8\in [0.90,1.03]$. Hence the effective sound speed
varies very little in function of 
${\tilde \Omega}_{m}$.
Alternatively, considering the $\Lambda$CDM theoretical value of
$\gamma$ ($\equiv 6/11$) and minimizing with respect to $\Omega_{m0}$
we find ${\tilde \Omega}_{m}=0.243\pm 0.034$ (see also \cite{Nes08})
with $\chi^{2}_{min}/dof \simeq 7.37/8$.

Our best-fit $\gamma$ value is in agreement within 
$1\sigma$ ($\Delta \chi_{1\sigma}^{2} \simeq 1$) uncertainty with 
the theoretically predicted value of $\gamma_{\Lambda} \simeq 6/11$
(see the cross in Fig. 2). 
Finally, in Fig. 3, we plot the measured $A_{obs}(z)$
with the estimated growth rate
function, $A(z)=f(z)\sigma_{8}(z)$
(see solid line).

We would like to finish this section with the following observation: 
if the particles are created proportional
to the DM density (see, for instance \cite{LG92,LA99,ZIM00}) then 
at each point the growth of matter density
perturbations will speed up with respect to an homogeneous
particle creation rate, changing the growth factor and thus it 
could potentially affect the observational constraints. However, in our case
we do not face such a problem because the corresponding 
particle creation rate term ($\Gamma_{dm}{3H}=(H_{f}/H)^{2}$) 
in the matter dominated era becomes $\Gamma_{dm}/3H \propto 1/\rho$.
Notice that in order to obtain the latter relation we use Eq.(\ref{friedr}).

\section{Scalar field description}

Matter creation models constitute a possible way to 
explain the cosmic acceleration
without the introduction of a dark energy component. However, it is 
sometimes desirable to represent the cosmic evolution 
in a field theoretical language, i.e., in terms 
of the dynamics of an ordinary scalar field ($\phi$). In a point of 
fact, all the dynamical stages discussed here can be described 
through a simple scalar field model (for a similar analysis 
see \cite{ZIM00}).

To begin with, let us replace $\rho$
and $p_{tot}=p+p_{c}$ in Eqs. (\ref{friedr}) and (\ref{friedp}) by
corresponding scalar field expressions
\begin{equation}
\label{scal}
\rho \rightarrow \rho_{\phi} = \frac{\dot{\phi}^{2}}{2} + V(\phi), \;\;\;\;\;\;
p_{tot} \rightarrow p_{\phi} =\frac{\dot{\phi}^{2}}{2} - V(\phi) \;.
\end{equation}
Inserting the latter into the Friedmann's equations we can separate the scalar field
contributions and express them in terms of $H$ and $\dot{H}$, i.e.,
\begin{equation}
\dot{\phi}^{2} =-\frac{1}{4\pi G}\dot{H} \;,
\label{ff3}
\end{equation}
\begin{equation}
\label{Vz}
V=\frac{3H^{2}}{8\pi G}\left( 1+\frac{\dot{H}}{3H^{2}}\right)=
\frac{3H^{2}}{8\pi G}\left( 1+\frac{aH^{'}}{3H}\right) \;,
\end{equation}
where $\dot{H}=aHH^{'}$ and prime here denotes 
derivative with respect to the scale factor.
Now, considering that $dt=da/aH$, Eq. (\ref{ff3}) can be integrated
to give
\begin{equation}
\label{ppz}
\phi=\int \left( -\frac{\dot{H}}{4\pi G}\right)^{1/2} dt =
\frac{1}{\sqrt{4\pi G}}\int \left(-\frac{H^{'}}{aH}\right)^{1/2}da\;.
\end{equation}

\subsection{Early de Sitter - radiation: Deflationary stage}

In this case the Hubble function is given by Eq. (\ref{solH}). 
Obviously we can 
integrate Eq. (\ref{ppz}), in the interval $[0,a]$ to obtain
\begin{eqnarray}\label{ppz1}\nonumber
\phi(a) & = & \frac{1}{\sqrt{2\pi G}}\;{\rm sinh}^{-1}\left( \sqrt{D}a\right),\\
& = & \frac{1}{\sqrt{2\pi G}}{\ln}\left( \sqrt{D}a+\sqrt{Da^{2}+1}\right)\;.
\end{eqnarray}
Note that at the time of 
inflation ($a=a_{*}$) the corresponding scalar field is
\begin{equation}
\label{ppz1}
\phi_{*}=
\frac{1}{\sqrt{2\pi G}}\;{\rm ln}\left( \lambda+\sqrt{\lambda^{2}+1}\right)\;.
\end{equation}   
Also after some simple algrebra, the potential energy becomes
\begin{equation}
V(a)=\frac{H^{2}_{I}}{8\pi G}\;\frac{3+Da^{2}}{(1+Da^{2})^{2}},
\end{equation} 
or 
\begin{equation}
V(\phi)=\frac{H^{2}_{I}}{8\pi G}\;
\frac{3+{\rm sinh}^{2}(\sqrt{2\pi G} \;\phi)} 
{[1+{\rm sinh}^{2}(\sqrt{2\pi G}\; \phi)]^{2}}  \;.
\end{equation}
In the context of slow-roll
approximation, one can prove that the density fluctuations are of
the form $\delta_{H} \sim H^{2} / \dot{\phi}^2 \sim 10^{-5}$ 
(see Peacock, {\it Cosmological Physics} 1999). 
Using Eqs. (\ref{solH}) and (\ref{ff3})
the function $H^{2} / \dot{\phi}^2 $ becomes 
\begin{equation}
\frac{H^{2}}{\dot{\phi}^2}=-4\pi G\frac{H^{2}}{\dot{H}}=
-4\pi G\frac{H}{a H^{'}}=2\pi G\frac{(1+Da^{2})}{Da^{2}} \;.
\end{equation}
At the epoch of inflation ($a=a_{*}$) we get 
\begin{equation}
\frac{H^{2}}{\dot{\phi}^2}(a_{*})=2\pi G\frac{(1+\lambda^{2})}{\lambda^{2}} \;.
\end{equation}
Inserting the latter into the density fluctuation constrain 
we obtain 
\begin{equation}
\frac{1}{\lambda^{2}}\sim \frac{\delta_{H}}{2\pi G}-1 \sim 
\frac{\delta_{H}}{2\pi \;l^{2}_{pl}} \;,
\end{equation}
or 
\begin{equation}
\lambda\sim \sqrt{\frac{2\pi}{\delta_{H}}} \;l_{pl}\;,
\end{equation}
where $l_{pl} \simeq \sqrt{G}$ is the Planck length in units 
$\hbar=c\equiv 1$.

\subsection{Accelerating dark matter stage}

In this case the late evolution of the Hubble function is given by
Eq. (\ref{Hz}). 
Now, inserting Eq. (\ref{Hz}) as well as its
derivative $(H^{'})$ into Eq. (\ref{ppz})
and integrating we obtain:
\begin{equation}\label{phia2}
\phi(a) = A\ln \left[\frac{\sqrt{1-{\tilde \Omega}_{\Lambda} + {\tilde \Omega}_{\Lambda} a^{3}}
- \sqrt{1-{\tilde \Omega}_{\Lambda}}} {\sqrt{1-{\tilde \Omega}_{\Lambda} +
 {{\tilde \Omega}_{\Lambda} a^{3}} + \sqrt{1-{\tilde \Omega}_{\Lambda}}} }\right]\;,
\end{equation}
where $A = 1/{\sqrt{24\pi G (1-{\tilde \Omega}_{\Lambda})}}$. 
From Eqs. (\ref{Hz}) and (\ref{Vz}) it is possible to show that
\begin{equation}\label{vdea}
V(a) = \frac{H_0^{2}}{8\pi G} \left[{\tilde \Omega}_{\Lambda} - 
\frac{1+2{\tilde \Omega}_{\Lambda}}{2}a^{-3}\right]\;.
\end{equation}
Finally, combining Eqs. (\ref{phia2}) and (\ref{vdea}) we find
\begin{equation}\label{vphif}
V(\phi) =  B + C\cosh(\omega \phi)\;,
\end{equation}
where $B = H_0^{2}( 5{\tilde \Omega}_{\Lambda} + 
{\tilde \Omega}_{\Lambda}^{2})/32 \pi G(1 -{\tilde \Omega}_{\Lambda})$,
$C = - H_0^{2}(1+2{\tilde \Omega}_{\Lambda})
{\tilde \Omega}_{\Lambda}/32\pi G(1 - {\tilde \Omega}_{\Lambda})$ and 
$\omega = \sqrt{24\pi G (1-{\tilde \Omega}_{\Lambda})}$. 


\section{Final Remarks}

A new cosmology based on the production of massless
particles (in the early de Sitter phase) and CDM particles (in the
transition to a late time de Sitter stage) has been discussed. The
same mechanism avoids the initial singularity, particle horizon and
the late time coincidence problem of the $\Lambda$CDM model has been
phenomenologically eliminated ($\Lambda \equiv 0$) because there is 
no dark energy in our accelerating scenario. In particular, this means 
that the dark energy component 
may be only a ``mirage'' (an effective description),  
since it can be mimicked (globally and locally) by the 
gravitationally induced particle production mechanism 
acting in the evolving Universe.

In this scenario, the standard cosmic phases - a  radiation era
followed by an Einstein-de Sitter evolution driven by
nonrelativistic matter until redshifts of the order of a few - are
not modified. However, the model has two extreme accelerating phases
(very early and late time de Sitter phases) powered by the same
mechanism (particle creation). Therefore, it sheds some light on a
possible connection among the different accelerating stages of the
universe. In particular our model, since $H_f^{2} = \tilde{\Omega}_{\Lambda}
H_0^{2}$, where $\tilde{\Omega}_{\Lambda}\sim 0.7$ and $H_{0}\simeq
1.5\times 10^{-42}GeV$, it sets the ratio of the primeval and late
time de Sitter scales to be $\rho_{I}/\rho_{f}=(H_I/H_f)^{2} \sim
10^{122}$. This large number is ultimately obtained through a 
combination of thermodynamics and quantum field theory in curved 
spacetimes in virtue to the association of the Hawking-Gibbons 
temperature ($T = H/2\pi$) to the early de Sitter phase.  Due to the 
maximal radiation production rate, $\Gamma_{r,max}=3H_I$, the  model 
deflates from an unstable de Sitter with initial 
expansion rate ($H = H_I$). It is exactly this Gibbons-Hawking 
connection (and the present value of $H_0$) that lead for the given 
number under the proviso that  the Universe starts with the Planck 
temperature. In principle, such a result in the present context 
has no correlation
with the so-called cosmological constant problem (in this 
connection see \cite{LT96}). In a forthcoming communication, it will 
be shown that the above ratio is preserved even if the initial de Sitter 
phase is powered by a production rate term 
proportional to $H^{n}$, $n \geq 2$.

As it appears, the cosmic history discussed here is semiclassically
complete. However, there is no guarantee that the initial de Sitter
configuration is not only the boundary condition of a true quantum
gravitational effect. In other words, the very early de Sitter phase
may be the result of a quantum fluctuation which is further
semi-classically  supported by the creation of massless particles
(in this connection see \cite{Pri2} and Refs. therein).

Naturally, the existence of an early isothermal de Sitter phase
suggests that thermal fluctuations (within the de Sitter event
horizon) may be the causal origin of the primeval seeds that will
form the galaxies. Such a possibility  and its consequences for the
structure formation problem deserves a closer investigation and is
clearly out of the scope of the present paper.

We stress that our model provides a natural solution to
the horizon problem and finally it connects smoothly the radiation
and the matter dominated eras, respectively. At late times it also
mimics perfectly the cosmic expansion history of the concordance
$\Lambda$CDM model. In this context,
we discuss the behavior of the linear matter perturbations $\delta_{m}$
on subhorizon scales for the CCDM model and the main results are shown in
Figs. 2 and 3. 

For completeness, we have also represented the evolution of our model in terms of the dynamics of an ordinary scalar field ($\phi$)
and derived analytically the scalar field potential for two regimes: (i) when the Universe
evolves from an early de Sitter to a Radiation phase and (ii) when it goes to the CDM phase to a
late time de Sitter stage. 

At present, we also know that a more complete version of the late
time evolution must be filled with CDM ($\sim$ 96\%) and baryons
($\sim$ 4\%), and, unlike $\Lambda$ cosmology, the baryon to dark
matter ratio is a redshift function \cite{LJO,BL10}. In particular,
this means that studies involving  the gas mass fraction may provide
a crucial test of our scenario, potentially, modifying our present
view of the dark sector. Some investigations along the above
discussed lines are in progress and will be published elsewhere.

\vspace{0.3cm} {\bf Acknowledgments:} The authors are grateful to G.
Steigman, A. C. C. Guimar\~aes, J. F. Jesus and  F. O. Oliveira  for
helpful discussions. JASL is partially supported by CNPq and FAPESP
under grants 304792/2003-9 and 04/13668-0,
respectively. SB wishes to thank the Dept. ECM of the
University of Barcelona for their hospitality, and the 
financial support from the
Spanish Ministry of Education, within the program of Estancias de
Profesores e Investigadores Extranjeros en Centros Espanoles (SAB2010-0118).
And FEMC is supported by FAPESP under grants 2011/13018-0.

\end{document}